\documentclass[12pt,preprint]{aastex}

\begin{document}

\title{On the Constancy of the Characteristic Mass of Young Stars}

\author{Bruce G. Elmegreen \affil{IBM Research Division, T.J. Watson
Research Center, P.O. Box 218, Yorktown Heights, NY 10598, USA,
bge@watson.ibm.com} }
\author{Ralf S.\ Klessen}\affil{Zentrum f{\"u}r Astronomie, Institut f{\"u}r
Theoretische Astrophysik, Ruprecht-Karls-Universit{\"a}t Heidelberg,
Albert-Ueberle-Str. 2, 69120 Heidelberg, Germany,
klessen@ita.uni-heidelberg.de } \and
\author{Christine D. Wilson} \affil{Department of Physics \& Astronomy,
McMaster University, Hamilton, Ontario L8S 4M1 Canada, wilson@physics.mcmaster.ca}

\begin{abstract}
The characteristic mass $M_c$ in the stellar initial mass function
(IMF) is about constant for most star-forming regions. Numerical
simulations consistently show a proportionality between $M_c$ and the
thermal Jeans mass $M_J$ at the time of cloud fragmentation, but no
models have explained how it can be the same in diverse conditions.
Here we show that $M_J$ depends weakly on density, temperature,
metallicity, and radiation field in three environments: the dense cores
where stars form, larger star-forming regions ranging from GMCs to
galactic disks, and the interiors of HII regions and super star
clusters. In dense cores, the quantity $T^{3/2}n^{-1/2}$ that appears
in $M_J$ scales with core density as $n^{0.25}$ or with radiation
density as $U^{0.1}$ at the density where dust and gas come into
thermal equilibrium. On larger scales, this quantity varies with
ambient density as $n^{-0.05}$ and ambient radiation field as
$U^{-0.033}$ when the Kennicutt-Schmidt law of star formation
determines $U(n)$. In super star clusters with ionization and
compression of pre-stellar globules, $M_J$ varies as the 0.13 power of
the cluster column density. These weak dependencies on $n,$ $U,$ and
column density imply that most environmental variations affect the
thermal Jeans mass by at most a factor of $\sim2$. Cosmological
increases in $M_J$, which have been suggested by observations, may be
explained if the star formation efficiency is systematically higher at
high redshift for a given density and pressure, if dust grains are
smaller at lower metallicity, and so hotter for a given radiation
field, or if small pre-stellar cores are more severely ionized in
extreme starburst conditions.
\end{abstract}


\section{Introduction}

The stellar initial mass function (IMF) has three properties that
appear to be relatively robust in diverse environments: the power law
slope for masses between 1 and 10 M$_\odot$, originally measured by
Salpeter (1955), the lower mass limit for the power law and the broad
plateau below it before the brown dwarf regime (Miller \& Scalo 1979;
Scalo 1986; Rana 1987; Reid 1987), and the maximum mass of stars
(Weidner \& Kroupa 2004; Oey \& Clarke 2005; Koen 2006). This paper
considers the origin of the plateau, a range typically spanning a
factor of $\sim3$ on either side of a characteristic mass
$M_c\sim0.3\;M_\odot$ where the IMF becomes relatively flat on a $\log
N-\log M$ plot. Occasionally, an observed IMF has a small sub-peak
inside the plateau, but the mass of this peak could vary stochastically
from region to region and with different mass binnings, calibration
details, binary corrections, and completeness corrections. The
existence of a broad IMF plateau defining a characteristic mass is
robust, however. At lower mass, the IMF drops into the brown dwarf
regime and at higher mass it drops into the stellar range for F, A, B
and O main sequence types. Comprehensive reviews of cluster and field
IMFs may be found in Scalo (1998), Kroupa (2002), and Chabrier (2003),
a review of starburst IMFs is in Elmegreen (2005), and a review of
galactic scale IMFs is in Elmegreen (2006). Recent reviews of the
theory of the IMF are in Mac~Low \& Klessen (2004), Bonnell, Larson \&
Zinnecker (2007), and Larson (2007).

Observations and possible explanations for the IMF plateau are
summarized in the next two sections. The observations suggest a
remarkable uniformity in the characteristic stellar mass in spite of a
wide range in environmental factors that should affect it. The
explanations are still incomplete because the issue of constant $M_c$
is not usually addressed. In section \ref{sect:3reasons}, we show that
the quantity $T^{3/2}n^{-1/2}$ that appears in the equation for the
thermal Jeans mass for temperature $T$ and number density $n$ should be
virtually independent of environment in three important cases: in dense
cores where individual stars form, in the general ISM where star
formation satisfies the Kennicutt (1998) relation, and in massive dense
clusters where OB stellar radiation determines the pressure and
temperature. In section \ref{sect:redshift}, observations for a
systematic increase in $M_c$ with redshift are reviewed, and
explanations are offered for why this might be the case. Finally, we
summarize our conclusions in section \ref{sect:conclusions}.

\section{Observations of the IMF Plateau: Uniformity in the Local Universe}
\label{sect:obs}

Observations of the mass range for IMF plateaus are collected in Table
1. Stochastic variations and systematic uncertainties in measured IMFs
make the plateau imprecise, but we can define it well enough for the
present purposes. As mentioned in the introduction, a typical IMF rises
somewhat monotonically with approximately the Salpeter power law slope
from high mass stars down to $0.5\;M_\odot$ or $1\;M_\odot$, and then
it either levels off on a log-log plot or rises more slowly down to
several tenths of a solar mass, at which point it decreases into the
brown dwarf range.  We consider the plateau to be the relatively flat
part of such an IMF on a log-log plot, extending for about a full width
at half maximum. The plateau for a log-normal approximation to the IMF
is about the same as the plateau for a piece-wise power law
approximation. For local clusters and the local field, the plateau
extends between $\sim0.1$ and $\sim1$ $M_\odot$. The observations in
Table 1 constrain the plateau in each IMF to within a factor of $\sim2$
in mass. Variations that are smaller than this are probably stochastic
and in any case will not be identified with physical variations in this
paper.

The characteristic mass $M_c$ is defined to be the mid-point of the
plateau, again on a log-mass scale. With the above typical limits, it
is $\log M_c/M_\odot \sim-0.5\pm0.5$ or $M_c\sim0.3\;M_\odot$. This
approximate value for the characteristic mass is observed in a
surprisingly wide variety of regions and over a wide range of star
formation epochs and rates. Among this diversity are included a high
latitude cloud with unusual abundances, Blanco 1 (Moraux et al. 2007),
and some of the most remote clusters in the outer Milky Way, Digel 1
North and South (Yasui et al. 2008), which have IMFs like that in
Orion. For Milky Way halo globular clusters, there is also a plateau
with similar $M_c$. Paresce \& De Marchi (2000) studied 12 globulars
and found that all of them have the same IMF shape even though they
represent a wide range of metallicities, distances from the galactic
center and plane, cluster radii and concentration ratios; there was
also no correlation between the IMF and the cluster disruption time. De
Marchi et al.\ (2000) noted that globular cluster IMFs resemble that in
the Pleiades. The only exceptions for globular clusters seem to be
those that have age-dependent mass segregation, which can severely
affect the present-day mass function (De Marchi, Paresce \& Portegies
Zwart 2004), and those that are highly dispersed by tidal forces, which
can have inverted present-day mass function (rising toward higher mass)
because of a preferential loss of low mass stars (de Marchi et al.\
1999; Koch et al.\ 2004; de Marchi, Pulone, \& Paresce 2006; De Marchi
\& Pulone 2007).

Distant globular clusters where the IMF is not observed directly should
have about the same value of $M_c$ as local globular clusters in order
to have remained bound after stellar evolution removed gas through
massive stellar winds and supernovae (de Grijs \& Parmentier 2007).
Globular clusters with higher initial $M_c$ would presumably have been
disrupted by such mass-loss processes, making $M_c\lesssim0.5\;M_\odot$
a selected value for survival. We do not know if $M_c$ was the same for
all globular clusters, considering this selection effect (Pfenniger,
private communication).

In the Galactic bulge, Holtzman et al.\ (1998) found an IMF that
rises with mass according to the Salpeter slope, $\sim-1.2$, down to
$\sim0.5\;M_\odot$, and then turns over sharply. Zoccali et al.\
(2000) traced the bulge IMF to lower mass and found about the same
slope as Holtzman et al.\ (1998) for $M>0.5\;M_\odot$ and a
shallower slope, $-0.33$, down to at least $0.16\;M_\odot$. This
implies that $M_c\sim0.3\;M_\odot$ in the bulge, similar to the
values seen in the globular clusters. The Galactic spheroid has an
IMF with a plateau peaking at $\sim0.6\;M_\odot$ (Gould et al.\
1997), slightly larger than the cluster plateau. The IMF in the
local dwarf Spheroidal Ursa Minor has been traced down to
$0.45\;M_\odot$ by Feltzing, Gilmore \& Wyse (1999) and found to be
indistinguishable from that of the globular cluster M92. The
observed mass range in Ursa Minor is too small to see both sides of
the plateau, but a significant increase in $M_c$ can be ruled out.

Chabrier (2003) compiled observations of IMFs for Milky Way disk and
globular clusters and for the bulge, using a slightly different
definition of $M_c$ and considering also binary star corrections. He
suggested $M_c$ is slightly larger for the older systems, with a shift
from $M_c\sim0.08\;M_\odot$ for the disk to $\sim0.2\;M_\odot$ for
globulars and the bulge if the binary fraction in the bulge is
significantly smaller than in the disk.  These values are uncertain and
the proposed shift is small, considering the large differences in star
formation rates and luminosity densities when these systems formed. We
return to these observations in Section \ref{sect:redshift}.

IMF variations under extreme environmental conditions such as
circum-nuclear starbursts are still uncertain. Elmegreen (2005)
noted that most previous claims of top-heavy or high-$M_c$ IMFs in
starbursts have been overturned by recent data. Bastian \& Goodwin
(2006) suggested that even the recent discussions about top heavy
IMFs in some super star clusters are premature as these clusters
appear unrelaxed and their masses uncertain.  In the Milky Way, the
Arches cluster near the Galactic center was reported to have a large
$M_c$ (Yang et al. 2002; Stolte et al.\ 2005) but it is poorly
observed at low mass. The most recent observations by Kim et al.\
2006 are incomplete below 1.2 $M_\odot$.  R136 in 30 Dor was also
claimed to have a high $M_c\sim2$ $M_\odot$ (Sirianni et al.\ 2000),
but extinction variations could affect this (Andersen \& Zinnecker
2003; Andersen et al.\ 2005).

Most extragalactic regions are too far away to observe the IMF
plateau directly, but limits can be placed on its range.
Significantly higher $M_c$ will produce too red a stellar population
of giants without the corresponding main sequence stars after the
turnoff age reaches the stellar lifetime at the lower limit of the
power law (Charlot et al.\ 1993). Also, the oxygen abundance will be
too high compared to solar after the turnoff age if the plateau
shifts upward by a factor of a few (Wang \& Silk 1993).
Significantly lower plateaus would produce too high a mass-to-light
ratio in the disk.

The uniformity of $M_c$ for local or normal star formation is
difficult to understand considering the wide range of properties in
molecular clouds. The radiation field, density, and temperatures for
gas and dust are high in super star clusters and low in dispersed
regions like Taurus, $\rho$ Oph, and IC 348, yet all of these
regions have about the same $M_c$. Globular clusters, which
presumably formed as super star clusters, also have about this
$M_c$. Indeed, the possible explanations for the plateau, discussed
next, do not indicate why $M_c$ should be so constant.

\section{Explanations for the IMF Plateau}

There is no convincing explanation for the constancy of $M_c$ in these
highly diverse regions. Although there are several proposals for the
origin of $M_c$, its invariance under varying environmental conditions
is usually not addressed. These proposals include an accretion- or
coagulation-driven scale-up of the opacity-limited mass (Field \&
Saslaw 1965; Rees 1976; Yoshii \& Saio 1985; Murray \& Lin 1996; Bate,
Bonnell, \& Bromm 2003), the thermal Jeans mass at an inflection point
in the effective equation of state (Larson 2005; Jappsen et al.\ 2005;
see also Li et al.\ 2003), and the initial thermal Jeans mass in a
marginally unstable cloud (Larson 1978; Tohline 1980; Klessen et al.\
1998; Klessen \& Burkert 2000; 2001). There are other models which
relate the IMF to the properties of interstellar turbulence (Larson
1981, Fleck 1982; Elmegreen 1993; Padoan 1995; Padoan et al.\ 1997;
Klessen 2001; Padoan \& Nordlund 2002). Also stellar winds may play
some role in self-limiting the stellar mass to a characteristic value
(Larson 1982; Shu, Adams, \& Lizano 1987; Nakano, Hasegawa, \& Norman
1995; Adams \& Fatuzzo 1996).

The thermal Jeans mass for an isothermal cloud is
\begin{equation}M_J=
\left(kT/Gm_{\rm H_2}\right)^{1.5}\rho^{-0.5}= 0.9\left(T/10\;{\rm
K}\right)^{1.5}\left(n_{\rm H_2}/10^4\;{\rm
cm}^{-3}\right)^{-0.5}\;M_\odot,\label{eq:mj}\end{equation} where $k$
and $G$ are the Boltzmann and gravitational constants, respectively,
and where mass density $\rho$ and number density $n_{\rm H_2}$ are
related via the H$_2$ mass, $m_{\rm H_2}$. The Bonner-Ebert critical
mass for an external pressure $P_{ext}$ is
\begin{equation}M_{BE}=1.18\left(kT/m_{\rm
H_2}\right)^2\left(G^{3}P_{ext}\right)^{-0.5}.\end{equation} If we use
a parameter $\alpha$ to connect the internal and external pressures of
the sphere, $P_{ext}=\alpha\rho kT/m_{\rm H_2}$, then
$M_{BE}=1.18\alpha^{-1/2} M_J$. Magnetic fields are generally ignored
for applications of these expressions to star formation because the
field eventually diffuses out. The minimum critical mass for stability
is determined by $M_J$ or $M_{BE}$.

Without additional assumptions, these explanations for $M_c$ are
difficult to reconcile with the diversity of star formation
conditions. The opacity-limited mass, in which opacity is limited by
dust, should vary inversely with metallicity (Masunaga \& Inutsuka
1999), so this explanation has difficulty with the similarity in
$M_c$ for modern disk clusters and halo globular clusters, which
have $\sim10-100$ times lower metallicity.  A wind-limited mass is
inconsistent with the apparent similarity between stellar mass
functions and protostellar core mass functions (for recent
observations and other references, see Alves et al.\ 2007; Ikeda et
al.\ 2007; Nutter \& Ward-Thompson 2007; Li et al.\ 2007; Walsh et
al. 2007; Massi 2007; see also Clark, Klessen, \& Bonnell 2007 for a critical
note on that issue).

The appearance of a characteristic mass from an inflection point in
the effective equation of state (EOS) has been demonstrated
numerically for both the first and second generation of stars in the
Universe (Bromm et al. 2002; Clark, Glover, \& Klessen 2008) and present day
conditions (Jappsen et al.\ 2005; Bonnell et al.\ 2006). For star
formation in the solar neighborhood, Larson (1985, 2005) suggested
that gas and grains couple thermally to $T\sim8$K at a density of
around $10^{-19}$ g cm$^{-3}$ ($n_{\rm H_2}=2.5\times10^4$
cm$^{-3}$), giving $M_J\sim2\;M_\odot$. According to the model, the
gas heats up with decreasing density below this coupling density,
and it heats up with increasing density above the coupling density
by equilibrating with a higher grain temperature. As a result, the
effective adiabatic index is less than 1 at the beginning of the
collapse, leading to fragmentation, and larger than 1 above the
coupling density, leading to little fragmentation. The Jeans mass at
the inflection point then ends up as the characteristic mass for
star formation, $M_c$. Whitworth, Boffin \& Francis (1998) also
suggested that $M_c\sim M_J$ at the gas-dust coupling point. In
Larson's model, an increase in temperature with density also follows
from the input of collapse energy. Masunaga \& Inutsuka (2000)
showed this collapse temperature increase, starting at a temperature
of $5\,$K and a density of $10^{-17.5}$ g cm$^{-3}$ ($n_{\rm
H_2}=8\times10^5$ cm$^{-3}$), which gives $M_J\sim 0.035\;M_\odot$,
considerably lower than $M_c$. However, at the gas-dust coupling
density in Masunaga \& Inutsuka (2000), the temperature still
decreased with increasing density because background starlight
radiation was increasingly excluded. Thus there is no inflection
point there. Observations of pre-stellar dense cores confirm this
temperature trend, showing decreasing temperatures with increasing
densities up to at least $10^6$ cm$^{-3}$ (Crapsi et al 2007). Thus
the EOS inflection point may occur too late in the collapse to set
$M_c\sim M_J$, or it may not occur at all.

The other explanation for the characteristic mass, found in numerical
simulations of various types (Klessen \& Burkert 2000, 2001; Clark \&
Bonnell 2005; Bate \& Bonnell 2005; Martel et al.\ 2006), is that it
scales with the thermal Jeans mass $M_J$ at the onset of collapse. In a
statistical sense, the system retains knowledge of its initial average
properties during gravitational contraction and the build up of dense
stellar clusters. We note, however, that these calculations are usually
done with an isothermal EOS and do not take compressional heating or
stellar feedback into account. They also describe evolutionary stages
before the collapse energy significantly increases the core
temperature, and thus correspond to a lower density and perhaps a
higher temperature, giving a higher $M_J$ than in the grain-gas
coupling theory. According to Clark \& Bonnell (2005), turbulent
fragmentation makes unbound cores ($M<M_J$) that coalesce and become
gravitationally bound when they reach $M_J$. Then they fragment into
lower-mass stars in a regular way, preserving the initial sensitivity
to $M_J$. The coagulation process preserves the clump densities and
$M_J$ until the fragments become gravitationally bound, at which point
their densities increase and they fragment gravitationally into stars
of lower mass. The sensitivity to $M_J$ in the pre-collapse cloud
remains in the stars that form.

For both the EOS condition and the initial-$M_J$ condition, the
environmental dependence of $M_c$ should be examined.  An important
consideration is the dependence of dust temperature $T_d$ on the local
density of the star-forming clump, which determines $M_J$ at grain-gas
coupling. If $T_d\propto n^{1/3}$, then $M_J$ is about constant.
Whether or not the EOS has an inflection point there does not matter,
because as long as $M_J$ is roughly constant near the beginning of the
collapse, the IMF should have a constant $M_c$ at the end of the
collapse, according to Clark \& Bonnell (2005) and others mentioned in
the previous paragraph.  A related dependency is that of gas and dust
temperatures on the average density in the star-forming cloud or the
average density in a region of the galaxy. These are more general
considerations that should determine how $M_J$ varies on larger scales.
The value of $M_J$ in a dense super star cluster should also be
examined because the pressure and temperature in a neutral prestellar
clump depend mostly on the ionizing radiation field. These three
environmental dependencies are discussed in the next section.
Remarkably, they all give about the same $M_J$ and show very little
sensitivity to environment.

\section{Three Reasons for a Constant $M_J$ in Various Environments}
\label{sect:3reasons}
\subsection{High Density Gas in Star Forming Regions}
\label{sect:highdensity}

The thermal Jeans mass depends on $T^{3/2}n^{-1/2}$ in a molecular
cloud. In general the temperature $T$ and density $n$ should be
independent, making $M_J$ vary with cloud conditions. At the high
densities where stars form, however, grain-gas coupling is an
important source of heating and cooling, making the gas and grain
temperatures comparable. The primary source of gas heating at
moderate to high density is the warm dust heated by starlight,
provided the magnetic diffusion rate is not greatly elevated by
temporary compression. The energy equation for the gas then has the
heating rate from grain collisions equal the cooling rate from line
emission.

Molecular line emission cooling has been studied by Neufeld et al.
(1995), who show in their Figure 2 the total cooling rate per H$_2$
molecule at various densities and temperatures. Their cooling rates per
molecule are approximately constant at each temperature for densities
ranging between $10^4$ cm$^{-3}$ and $10^8$ cm$^{-3}$. These rates vary
from $\log\left(\Lambda/n\right)\sim-26.5$ (in ergs s$^{-1}$) at
$T=10\,$K to $-25.8$, $-25.0$, and $-24.0$ in these units at $T=20\,$K,
40$\,$K, and 100$\,$K, respectively. On average, this is approximately
linear on a log-log plot, with a change in $\log\left(\Lambda/n\right)$
by 2.5 orders of magnitude for a change in $T$ by 1 order of magnitude.
Thus the cooling rate is approximately $\Lambda=10^{-26.5}n_{\rm
H_2}\left(T/10\,\mathrm{K}\right)^{2.5}$ erg s$^{-1}$ where $n_{\rm
H_2}$ is the number density of H$_2$ molecules in cm$^{-3}$. Setting
this equal to the gas heating rate from dust with number density $n_d$,
temperature $T_d$ and characteristic radius $a$ (cf. Hollenbach 1989;
Tielens 2005, Eq. 3.27), we obtain
\begin{equation}
\alpha_a \pi a^2 c_{th} n_{\rm H_2}n_d 2k(T_d-T)=10^{-26.5}n_{\rm
H_2}\left(T/10\,\mathrm{K}\right)^{2.5}
\end{equation}
for cgs units. We can now solve for the quantity appearing in $M_J$,
\begin{equation}
T^{3/2} n^{-1/2}=10^{21.75}\left(\alpha_a \pi a^2D2k\Delta T
\left[8k/\pi m_{\rm H_2}\right]^{1/2}\right)^{3/4}n^{1/4}\label{eq:t32}
.\end{equation} In these equations, $c_{th}=\left(8kT/\pi m_{\rm
H_2}\right)^{1/2}$ is the three-dimensional thermal speed of collisions
between gas and dust, $\alpha_a= 0.15$ to 1 measures how well the gas
atom thermalizes while on the grain (Tielens 2005), $\Delta T=T_d-T$,
and $D=n_d/n_{\rm H_2}$. The temperature difference $\Delta T$ is
assumed to be small and can be defined as independent of $T$ at the
coupling density.

Evidently, the Jeans quantity, $T^{3/2}n^{-1/2}$, depends only
weakly on density at the grain-gas coupling point, as $n^{1/4}$. It
depends weakly on grain properties too, because $a^2D$ averaged over
a grain size distribution proportional to $a^{-3.5}$ (Mathis et al.\
1977) depends mostly on the smallest grains and is therefore
approximately independent of the dust-to-gas mass ratio and
metallicity, which depends on the largest grains (see Eq.
\ref{eq:13} below). The metallicity dependence comes mostly from the
gas cooling rate $\Lambda$.  If we assume $\Lambda\propto Z^\alpha$
then $M_J\propto \left(a^2D/Z^\alpha\right)^{3/4}$ for metallicity
$Z$. If the main coolants are optically thick, then the exponent
$\alpha$ should be small and the overall metallicity dependence
should be weak, especially if $a^2D$ decreases a little at lower $Z$
along with $\Lambda$.

To emphasize the weakness of the $n^{1/4}$ density dependence for
$M_J$, we note that $n\propto T^2$ at grain-gas coupling, and for
typical grain properties, $T\propto U^{0.2}$ with radiation field
$U$ (Tielens 2005; eq. 5.44). Thus $n\propto U^{0.4}$ and so the
$n^{1/4}$ dependence in the Jeans quantity is $\propto U^{0.1}$. The
Jeans mass at grain-gas coupling hardly depends on the most
important environmental variable, the radiation field. For example,
if the radiation field in the immediate vicinity of a pre-stellar
clump increases by three orders of magnitude, the dust and gas
temperatures at coupling both increase by a factor of $\sim4$, the
density at coupling increases by a factor of 16, but the Jeans mass
increases by only a factor of 2.

An evaluation of $M_J$ from equation \ref{eq:mj} is somewhat uncertain
as it requires several assumptions about grain properties. The most
important result for this paper is the extreme insensitivity of $M_J$
to environment at the grain-coupling density. Still, the reader may be
interested in the value of $M_J$ derived in this way, so we make an
attempt here.  We assume that the size distribution of grains is
$dn_g/da=n_a a^{-3.5}$ (Mathis et al. 1977), the maximum grain size is
$a_{max}=2\;\mu$m, the minimum grain size is $a_{min}=0.005\;\mu$m, the
grain specific density is $\rho_{g}=1$ g cm$^{-3}$, and the total
dust-to-gas mass ratio is 0.01. The grains are also assumed to be
spherical. This is enough to derive the mean quantity $\langle a^2 D
\rangle=0.01 \left(m_{H_2}/\rho_g\right)\times\left(3/4\pi\right)\times
\left(a_{max}a_{min}\right)^{-1/2} \sim9.6\times10^{-22}$ cm$^{2}$ that
appears in equation \ref{eq:t32}. If we also take thermalization
coefficient $\alpha_a=0.5$, gas-grain temperature difference $\Delta
T=2$K, and density $n\sim10^5$ cm$^{-3}$ at coupling (which corresponds
to a solution of eq. \ref{eq:t32} for $T=9$K), then
$M_J=0.24\;M_\odot$, which is a reasonable value for the characteristic
mass.

The increasing dependence of $M_J$ on $n^{1/4}$ is too weak to imply
that low density regions of GMCs should produce more low mass stars.
The relative proportion of high and low mass stars depends more on the
overall shape of the IMF, which is not discussed in this paper.

\subsection{Ambient Gas in Star Forming Regions}
\label{sect:ambient}

The ambient gas density in star forming regions varies from a low
galactic average of $\sim1$ cm$^{-3}$ or less in spiral galaxy disks to
a high molecular cloud average of $\sim10^4$ cm$^{-3}$ or more for
giant molecular cloud complexes (GMC) and inner-disk starburst regions.
The characteristic mass in the IMF hardly varies throughout these
regions so there has to be some kind of regulation to keep
$T^{3/2}n^{-1/2}$ constant if $M_c\propto M_J$. In the previous
section, we discussed only the dense cores where small stellar groups
and binary stars form, at densities close to the value where dust and
gas become thermally coupled. Here we discuss the lower-densities
surrounding these cores, down to the ambient density in the galactic
interstellar medium.

The most important regulator for star formation on scales that range
from galactic disks to molecular clouds is the dynamical nature of
the star formation process, which gives a rate per unit volume
proportional to $n^{1.5}$ (e.g., Elmegreen 2002). This could be the
origin of the Kennicutt (1998) law of star formation, sometimes
called the Kennicutt-Schmidt (KS) law. The dynamical rate enters
into the  $n^{0.5}$ dependence, and the other factor of $n$ is from
the mass per unit volume of available gas. There may be other
dependencies on environmental parameters in this relation, such as
the Mach number, mean magnetic field strength, or external radiation
(V\'azquez-Semadeni et al.\ 2003; Schaye 2004; Krumholz \& McKee
2005), but these dependencies are difficult to observe directly. The
KS law relating the star formation rate per unit area to the mass
surface density of gas per unit area has been measured on a wide
range of scales, from whole galaxies (Martin \& Kennicutt 2001) to
pieces of galaxies (Kennicutt et al. 2007). There are indications
that either the coefficient or the power may vary with the local
rate of shear (Vorobyov 2003; Luna et al.\ 2006), and there are
uncertainties regarding the CO to H$_2$ conversion and other
calibrations (Boissier et al. 2003), but the basic power law form
seems to be robust. Even models that assume a local star formation
rate proportional to the first power of the molecular density, and
then calculate the molecular fraction in various environments,
recover the average KS relation with power 1.5 that we use here
(Robertson \& Kravtsov 2007).

The KS law is based on observations of column density and here we
consider volume density. This is a reasonable conversion if the gaseous
scale height is about constant with radius in a galaxy, which is
approximately true in our own Galaxy (e.g., Sanders, Solomon \&
Scoville 1984). It is also reasonable if the photon mean free path in
the near-UV is comparable to the disk thickness, which is also
approximately true. The heating rate per grain depends on the mean
intensity of radiation, which is the product of the volume emissivity
and the photon mean free path averaged over all directions. The volume
emissivity is about the star formation rate per unit area divided by
the disk thickness, so if the mean free path is comparable to the disk
thickness, then the mean intensity of radiation is proportional to the
star formation rate per unit area, as assumed here.

On scales smaller than the galactic scale height, the $\rho^{1.5}$
dependence for the star formation rate is still a reasonable assumption
considering the dynamical nature of the processes involved. The use of
a local radiation field proportional to this rate is also reasonable
because star formation generally dominates the background heat sources.
We discuss below how a time dependence might change the results.

We now ask how the dust temperature varies with density on the
scales where the KS law operates. The dust temperature for general
dust composition varies with grain size $a$ and ambient radiation
field $G_0$ as (Tielens 2005, eq. 5.44)
\begin{equation} T_d=33.5\left(a/1\,\mu\mathrm{m}\right)^{-0.2}\left(
G_0/10^4\right)^{0.2} \;{\rm K}\label{eq:dust}\end{equation} where
$G_0$ is measured in units of the Habing field, $1.6\times10^{-3}$
erg cm$^{-2}$ s$^{-1}$. This relation assumes the Planck mean
efficiency of grain absorption scales inversely with wavelength (a
wavelength-squared dependence would change the exponents to $1/6$).

In a star-forming region the radiation field $G_0$ will be proportional
to the star formation rate, which is proportional to $n^{1.5}$ by the
KS law. Thus $T_d\propto n^{0.3}$. At the locally high densities of
star formation, where the dust and gas temperatures are comparable, the
environmentally-dependent Jeans mass quantity, $T^{3/2}n^{-1/2}$, is
therefore dependent on ambient density only weakly, as $n^{-0.05}$.
Using the KS law again, this corresponds to an $M_J\propto
G_0^{-0.033}$ dependence, which is also very weak. For an increase in
star formation rate and radiation field by 3 orders of magnitude, $M_J$
decreases by only a factor of 1.3 when the gas and dust temperatures
are comparable, even at densities that are not the thermal coupling
density. At thermal coupling, $M_J$ increases by a factor of 2, from
the previous subsection.  The necessary precision to observe these
variations in M$_J$ is not available yet, explaining why the observed
characteristic mass for star formation is so constant, spanning the
range from ambient field regions like Taurus to intense starburst
regions like interacting galaxies.

The cores where individual stars form are much denser than the ambient
ISM discussed above, so the constancy of $M_J$ on large scales may seem
irrelevant to the IMF. However, if star formation operates at the
dynamical rate for a wide range of scales, as seems to be the case
(Elmegreen \& Efremov 1996; Elmegreen 2007), then the star formation
rate should be proportional to $n^{1.5}$ for a wide range of densities,
possibly even including the inner cores of dense clusters, where the
IMF is determined. Similarly the radiation field should scale with the
local star formation rate for a range of densities, considering time
and space averages of this field. In this case the constancy of $M_J$
from the KS relation would also apply to cloud cores. An interesting
exception should occur at the onset of star formation, because the
density can be high before the radiation field or temperature are high,
lowering $M_J$ below the characteristic value.  This would seem to be a
problem for inactive regions like the Pipe Nebula (Alves et al. 2007)
that have cold and dense prestellar cores with a characteristic mass a
factor of $\sim3$ above $M_c$ for stars. However, these cores began
forming from lower density material in the Pipe Nebula, and at that
time, the discussions in Clark \& Bonnell (2005) apply: $M_J$ in the
turbulent medium has an important influence on $M_c$ even before
self-gravity becomes important on small scales. Indeed, the Pipe Nebula
cores near the mass function turnover are not gravitationally bound
(Lada et al.\ 2008) so the $M_J$ argument of section
\ref{sect:highdensity} does not apply to them yet. The difficult
question is not {\it whether} $M_J$ is constant, which appears to be
the case in a variety of situations, but {\it when} $M_c\sim M_J$ is
established in the life cycle of a molecular cloud.

The value of $M_J$ for average GMC and lower-density conditions depends
on the radiation field that is expected for a star formation rate given
by the KS law. The star formation rate ${\dot M}$ is related to the
total IR luminosity as ${\dot
M}=2\times10^{-10}\left(L_{IR}/L_\odot\right)\; M_\odot$ yr$^{-1}$
(Kennicutt 1998).  This equation assumes that all of the radiation from
young stars comes out in the IR, so this luminosity is the total from
the star-forming region. The Kennicutt (1998) relation in which the
star formation rate per unit area depends on the $\sim1.5$ power of the
column density of gas was converted by Elmegreen (2002) into a relation
between the star formation rate per unit volume and the 1.5 power of
the volume density. It assumes that the average ISM density is
comparable to the threshold tidal density in the galaxy, which is
approximately true everywhere, and that the disk has an exponential
light profile with a flat rotation curve. This gives the result ${\dot
M}/{\rm Volume} \sim 0.012\rho\left(G\rho\right)^{1/2}$ in cgs units.
Thus the luminosity density from star formation is
\begin{equation}
L/{\rm Volume}=3.8\times10^{15}\rho\left(G\rho\right)^{1/2}
\end{equation}
in cgs units. The volume emissivity is this luminosity density divided
by $4\pi$, and the radiation field $G_0$ is the volume emissivity times
the path length. For a path length of $L=1$ kpc, which is about one
optical depth, $G_0=1.9\times10^{-3}\,n_{\rm H_2}^{1.5}$ erg\,
cm$^{-2}\,$s$^{-1}$, where $n_{\rm H_2}$ is the ambient density
measured in molecules per cubic centimeter for convenience in scaling
to higher densities.  The Habing radiation field is $1.6\times10^{-3}$
erg cm$^{-2}$ s$^{-1}$, so this value from the star formation rate is
1.2 times the Habing field for a KS relation at unit molecular density.
The dust temperature from equation \ref{eq:dust} is therefore 5.5$\,$K
for $a/1\;\mu{\rm m}=1$. In general, the dust temperature from star
formation heating alone is
\begin{equation}
T_d=5.5\left(a/1\;\mu{\rm
m}\right)^{-0.2}\left(L/{1\,\mathrm{kpc}}\right)^{0.2}n_{{\rm H}_2}^{0.3}
\end{equation} based on the KS law with the above assumptions. For
ambient radiation there would be other sources in addition to star
formation, so this temperature is low compared to the observed ambient
$T_d$. This equation is useful for scaling to higher star formation
densities and their correspondingly higher radiation fields.

At this point in the evaluation of $M_J$, we need to assume some
density where new star formation occurs; the density used in the
previous paragraph is the average in the region producing the
radiation, which is assumed to follow the KS law. We designate the
density in the radiating star-formation region $n_{rad}$, and the
density in the new star-forming region, where $M_J$ is to be
determined, $n_{sf}$. If new star formation occurs in a typical
molecular cloud core, $n_{sf}\sim10^4\,$cm$^{-3}$. Putting $T_d$ and
$n_{sf}$ into equation \ref{eq:mj}, we get
\begin{equation}
M_J=0.37\left(n_{rad}/1\;{\rm cm}^{-3}\right)^{0.45}
\left(n_{sf}/10^4\;{\rm cm}^{-3}\right)^{-0.5} \left(L/{1\,\mathrm{kpc}}\right)^{0.3}
\;M_\odot .
\end{equation}
If the radiation path length, $L$, is about constant and $n_{sf}\propto
n_{rad}$, we obtain the very weak scaling of $M_J$ with density
discussed earlier in this subsection. The radiation path length should
scale inversely with $n_{rad}$, however, as smaller regions of star
formation typically have higher densities. For Larson-law scaling or
for constant absorption on the line of sight, this would be an exact
inverse relation. The density of new star formation should also depend
on $n_{rad}$, as pressure from the HII region around the radiating
stars comes to equilibrium with the pressure in the new star-formation
region (see section \ref{sect:compressed} for more on this situation).
For this latter relation, we take an ionizing luminosity per unit
volume proportional to the star formation rate per unit volume, which
scales with $n_{rad}^{1.5}$ according to the KS law, and then the
density of the HII region should scale with the square root of this, as
$n_{rad}^{0.75}$. If we consider all of this, i.e., $L\propto
n_{rad}^{-1}$ and $n_{sf}\propto n_{rad}^{0.75}$, we get a Jeans mass
scaling as $M_J\propto n_{rad}^{-0.225}$. This is a weak dependence on
environmental density, but not as weak at the $M_J\propto n^{-0.05}$
relation derived at the beginning of this sub-section. For $n_{rad}$
varying by a factor of $10^3$ from the ambient medium to the average
density in a GMC, $M_J$ decreases by a factor of $\sim5$ in this
expression.

We can summarize the present results as follows. For star formation in
dense cores that are exposed to radiation and pressure from a
surrounding region that may range in density from the ambient value in
a galaxy to the average in a GMC, the temperature and local density
scale together in such a way that the thermal Jeans mass is
approximately constant and comparable to the characteristic mass in the
IMF. All that has been assumed is that the radiation density scales
with the surrounding star formation rate density, and that the star
formation rate scales with the gas density multiplied by the dynamical
rate, i.e., according to the Kennicutt-Schmidt law written in
three-dimensional form. If the path length for the accumulated flux of
this radiation is constant, and the local density for star formation
scales with the surrounding density, $n$, then $M_J$ scales extremely
weakly with density, as $n^{-0.05}$. If the path length scales
inversely with surrounding density, whether by Larson's laws for star
formation in GMCs or by a constancy of the absorption, and if the local
density for star formation scales with the surrounding ionization
density, then $M_J$ scales as $n^{-0.225}$. In both cases, the KS law
is a unifying factor that helps to preserve a nearly constant $M_J$ in
a wide range of environments.

\subsection{Compressed Globules in HII Regions}
\label{sect:compressed}

HII regions provide a very different environment for star formation
compared to the dark neutral clouds considered in the previous two
subsections. It is important to consider how $M_J$ varies for the most
extreme conditions when neutral clouds and cloud pieces inside and
adjacent to HII regions are heated, compressed, and ionized by strong
radiation fields.  Such conditions should apply to any cluster massive
enough to form O-type stars, including super star clusters and young
globular clusters. Inside and surrounding the cloud cores where these
clusters form, there should be neutral globules, proplyds, elephant
trunks, and other neutral material that is still forming stars. Each
neutral piece will be heated and compressed by the first massive stars
that form in the cluster, giving an $M_J$ specific to that region.
Before these first massive stars appear, the results of the previous
two subsections should apply.

We use the analysis of Bertoldi \& McKee (1990) to determine the
critically stable mass for gravitational collapse in a
pressure-confined globule with an ionized boundary. This mass, given in
their equation 5.5 for non-magnetic clouds, is analogous to $M_J$ or
$M_{BE}$ in the previous discussion and so presumably is related to the
characteristic mass in the IMF for this environment. We also use the
bolometric luminosities $L$ and Lyman continuum ionization rates $S$
for massive stars from Vacca, Garmany \& Shull (1996), which were
integrated over an IMF in Elmegreen (2007, Fig. 1) to give these
quantities as functions of the cluster mass, $M$. For clusters more
massive than $\sim10^3$ $M_\odot$, where O-type stars are present, the
integrals give nearly linear dependences between both $L$ and $S$ and
the cluster mass,
\begin{equation}L\sim10^6\left(M/10^3\;M_\odot\right)\;\;\;
\mbox{and}\;\;\;S_{49}\sim10\left(M/10^3\;M_\odot\right) \;,
\label{eq:m3}
\end{equation}
where $S_{49}$ is the ionization rate in units of $10^{49}$ s$^{-1}$.

The critically stable mass from equation 5.5 in Bertoldi \& McKee
(1990) is
\begin{equation}
M_{crit}=47.2\,c^{14/3}\left(\frac{S_{49}}{(R/{1\,\mathrm{pc}})^2}\right)^{-1/3}\phi
\;\;M_\odot
\end{equation}
where $\phi$ includes a combination of factors of order unity, and $R$
is the distance to the star with ionization rate $S_{49}$, in units of
parsecs. This equation considers the pressurized boundary of the
near-spherical globule where the pressure is determined by the
ionization front. The thermal speed in the neutral cloud is $c$, in km
s$^{-1}$ (based on $c^2=kT/m_{\rm H_2}$). This thermal speed depends on
the dust temperature in the neutral gas, which depends on the radiation
field $G_0$ according to equation (\ref{eq:dust}). Tielens (2005)
writes $G_0=2.1\times10^2
\left(L/10^4\;L_\odot\right)\,(R/{1\,\mathrm{pc}})^{-2}$. Thus
\begin{equation}
T_d\sim15.5\left({{a}\over{1\,\mu\mathrm{m}}}\right)^{-0.2}
\left({{L/10^4\;L_\odot}\over{(R/{1\,\mathrm{pc}})^2}}\right)^{0.2}\;
{\rm K}\; =\; 38.9\left({{a}\over{1\,\mu\mathrm{m}}}\right)^{-0.2}
\left({{M/10^3\;M_\odot}\over{(R/{1\,\mathrm{pc}})^2}} \right)^{0.2}
\;{\rm K}
\end{equation} where we have used equation (\ref{eq:m3}). Inserting
this dust temperature into the sound speed we get
\begin{equation}
c=0.36\left({{a}\over{1\,\mu\mathrm{m}}}\right)^{-0.1}
\left({{M/10^3\;M_\odot}\over{(R/{1\,\mathrm{pc}})^2}} \right)^{0.1}\;{\rm km
\;s}^{-1}.\end{equation} Thus the critical mass is
\begin{equation}
M_{crit}=0.2\left({{a}\over{1\,\mu\mathrm{m}}}\right)^{-0.47}\left({{M/10^3\;M_\odot}\over{(R/{1\,\mathrm{pc}})^2}} \right)^{0.13}
\;M_\odot . \label{eq:mc}
\end{equation}
This critical mass is similar to the observed characteristic mass of
star formation and is only weakly dependent on cluster mass $M$ and
size $R$. Note that for heating and compression inside a cluster core,
the mean distance to the ionizing source, $R$, is the cluster size as
written here.

A magnetic field in the globule will stabilize it, so the critical mass
for collapse will be larger than in equation (\ref{eq:mc}) when a field
is present (e.g., Bertoldi \& McKee 1990). This is true for all of the
critical masses derived in this paper. The field will eventually weaken
because of ambipolar diffusion, so the non-magnetic result is most
relevant for the final state.

Another consideration for HII regions is the complete ionization of
small globules before they get compressed. Such ionization would
deplete the low mass part of the IMF and shift the characteristic mass
to a higher value, regardless of $M_{crit}$. This is possible for the
smallest globules according to equation 4.5 in Bertoldi \& McKee
(1990), where the mass limit for complete ionization may be written,
\begin{equation}
M_{ionize}<0.0189\left({{n}\over{10^3\;{\rm cm}^{-3}}}\right)^{-5}
\left({{S_{49}}\over{(R/{1\,\mathrm{pc}})^2}}\right)^{3}\;\;M_\odot=
19\left({{n}\over{10^3\;{\rm cm}^{-3}}}\right)^{-5}
\left({{M/10^3\;M_\odot}\over{(R/{1\,\mathrm{pc}})^2}}\right)^{3}\;\;M_\odot
. \label{eq:ionize}\end{equation} This expression is not very useful
because of the strong dependences on the initial globule density, $n$,
and the cluster mass, $M$, and size, $R$. Still, the mass limit is
comparable to $M_c$ when $n\sim2.3\times10^3$ cm$^{-3}$, a reasonable
value for a piece of the GMC before ionization compression.

Once a neutral globule reaches pressure equilibrium with its ionized
boundary, it either collapses quickly or is stable according to whether
its mass is greater than or less than $M_{crit}$. If it is unstable, it
cannot be ionized significantly because the collapse time is much
faster than the ionization time. Using equation 4.10a in Bertoldi \&
McKee (1990) for the ionization time and converting $S_{49}$ and $c$
into cluster mass $M$ as before, we get the ionization time
\begin{equation}
t_{ionize}\sim10^6\left({{a}\over{1\,\mu\mathrm{m}}}\right)^{0.12}
\left({{M/10^3\;M_\odot}\over{(R/{1\,\mathrm{pc}})^2}}
\right)^{-0.32}\,
\left(\frac{M_{globule}}{1\,M_{\odot}}\right)^{0.4}\;{\rm yr}
\end{equation} for $M_{globule}$ in $M_\odot$. The collapse time of a
compressed globule is much smaller than this. Equation 3.33c in
Bertoldi \& McKee (1990) gives the globule pressure in terms of $c$ and
$S_{49}/R^2$. Converting these variables into cluster mass $M$ and
dividing by $kT$ gives
\begin{equation}
n_{compressed}=1\times10^8 \left({{a}\over{1\,\mu\mathrm{m}}}\right)^{0.45}\left({{M/10^3\;M_\odot}\over{(R/{1\,\mathrm{pc}})^2}}
\right)^{0.15} \,\left(\frac{M_{globule}}{1\,M_{\odot}}\right)^{-0.2}\;\;{\rm cm}^{-3}.\end{equation} The
corresponding dynamical time (within a factor of 2 of the collapse
time) is
\begin{equation}
t_{dyn}=\left(G\rho\right)^{-0.5}=6\times10^3 \left({{a}\over{1\,\mu\mathrm{m}}}\right)^{-0.225}\left({{M/10^3\;M_\odot}\over{(R/{1\,\mathrm{pc}})^2}}
\right)^{-0.075} \,\left(\frac{M_{globule}}{1\,M_{\odot}}\right)^{0.1}\;\;{\rm yr} .\end{equation} The
dynamical time is always much less than the evaporation time in
pressure equilibrium, so if the globule does not get ionized
immediately, before the implosion, it will collapse to a dense core
with a much higher pressure and greater self-shielding ability. Then
a star will likely form.

\section{Extreme Star Formation: Low Metals and High Redshifts}
\label{sect:redshift}

The above considerations apply to star formation in the local
universe. If we go to high redshift or very low metallicity the
situation changes. Detailed numerical simulations predict
zero-metallicity (Population III)  stars to be massive,  $M > 20 \:
{M_{\odot}}$ (Abel, Bryan, \& Norman 2002; Bromm, Coppi, \& Larson
2002; Yoshida et al.\ 2006; O'Shea \& Norman 2007).  The lack of
zero-metal stars in the Milky Way is consistent with this (see
review in Beers \& Christlieb 2005). Thus $M_c$ must have been
higher at extremely low metal abundance (Tumlinson 2006).

The critical abundance for the transition in $M_c$ is debated.
Extremely metal-poor subgiant stars in the Galactic halo have masses
below one solar mass (Christlieb et al.\ 2002, Beers \& Christlieb
2005). The most extreme of these stars have [Fe/H] $<10^{-5}$,
although carbon and oxygen are still relatively high,
$\sim\,10^{-3}$ times solar. This unusual abundance pattern could be
produced by pair-instability supernovae in Population III (Heger \&
Woosley 2002) or mass transfer from close binary companions (Ryan et
al.\ 2005, Komiya et al.\ 2007). There are hints for an increasing
binary fraction with decreasing metallicity for these stars
(Lucatello et al.\ 2005). Some models suggest that low-mass star
formation becomes possible once atomic fine-structure line cooling
from carbon and oxygen becomes effective (Bromm et al.\ 2001;
Frebel, Johnson, \& Bromm 2007), setting a value for the transition
metallicity ${\rm Z_{\rm crit}}$ at around $10^{-3.5} \: {\rm
Z_{\odot}}$. However, for cold initial conditions and
$n<10^3\,$cm$^{-3}$, H$_2$ is the dominant coolant, suggesting that
the transition is determined by other physical processes such as
dust formation (Jappsen et al.\ 2007). An alternative view is that
low mass star formation is the result of dust-induced fragmentation
at high densities and late stages in protostellar collapse
(Schneider et al.\ 2002; Omukai et al.\ 2005; Schneider et al.\
2006). The transition metallicity is then in the range $10^{-6} \le
{\rm Z_{\rm crit}} \le 10^{-5}~{\rm Z_{\odot}}$ with the resulting
IMF plateau clearly falling below $1\: {M_{\odot}}$ at higher
abundances (Clark et al.\ 2008).

Observations that $M_c$ increases with redshift come from global
population studies of distant galaxies. These are not the hypothesized
Population III stars, but normal stars that are observed in present-day
mass functions. For example, van Dokkum (2008) studied mass-to-light
ratios and U-V colors for early-type galaxies in the range $0.02 < z <
0.83$. These are highly evolved galaxies and the mass-to-light ratio
depends mostly on the present day mass function near 1 $M_\odot$. The
best fit IMF for this sample is flat at 1 $M_\odot$, constrained by the
observed high rate of change of the mass-to-light ratio within this
redshift range.  This means that the IMF plateau has to include 1
$M_\odot$ within its range, unlike the local IMF which begins to
steepen into the Salpeter power law at this mass. Van Dokkum (2008)
also noted that a flat IMF at 1 $M_\odot$ is consistent with the
observed shallow rate of change of the Balmer absorption strength over
this redshift range. Constraints from cosmic background starlight and
the local luminosity density of galaxies also suggest an upward shift
in $M_c$. Fardal et al.\ (2007) found that a shallow power law slope is
not enough to simultaneously fit these two measurements, but a
``paunchy'' IMF with a peak at $\sim5$ $M_\odot$ is required. In a
third study, Dav\'e (2008) compared the star formation rate per unit
mass out to $z=2$ using three different surveys with that expected from
cosmological simulations that assume a fixed $M_c$. There was a clear
discrepancy that was explained by an increase in $M_c$ with redshift as
$0.5\left(1+z\right)^{2}\;M_\odot$, making $M_c$ larger by a factor of
$\sim9$ at $z=2$. For extreme metal-poor stars in the Milky Way, Komiya
et al.\ (2007) suggested that $M_c\sim5$ $M_\odot$ from the fractions
of these stars that are C-rich with and without s-process elements.
C-rich extreme metal-poor stars are probably surviving low-mass binary
members, so models of stellar evolution and binary mass transfer were
involved in their analysis.

Other studies of cosmological structure formation also suggest the IMF
shifts toward more massive stars, but most of these studies consider a
shallower slope in the power law part of the IMF above a few solar
masses, not a possible change in $M_c$, which could also be happening
(or happening instead of a shallower slope). These shallower slopes
include the starburst phases of massive elliptical galaxies (Pipino \&
Matteucci 2004; Nagashima et al.\ 2005b) and clusters of galaxies
(Renzini et al.\ 1993; Loewenstein \& Mushotsky 1996; Chiosi 2000;
Moretti, Portinari, \& Chiosi 2003; Tornatore et al.\ 2004; Romeo et
al.\ 2006; Portinari et al.\ 2004; Nagashima et al. 2005a). They also
include the Milky Way and M31 bulges, which appear to have had shallow
IMFs at intermediate to high mass because of their large [Fe/H]
abundances (Ballero et al.\ 2007). In the central parsec of the Galaxy,
the IMF for young stars may be shallow too (Nayakshin \& Sunyaev 2005;
Paumard et al.\ 2006; Maness et al 2007). Shallow in these studies
means an IMF slope of $\Gamma\sim=-1$ to $-1.1$, where the Salpeter
slope is $-1.35$. Recent theoretical models of flat IMFs are in
Klessen, Spaans, \& Jappsen (2007).

Evidently, the observations suggest that $M_c$ was higher at a redshift
of $\sim2$, perhaps by a factor of $\sim10$, even for stars that are
not zero-metallicity. What could have caused this increase if $M_c$ is
as independent of environment as the previous sections suggest? To
attempt to answer this question, we consider how $M_c$ might have
increased in the early universe for each of the three models in section
\ref{sect:3reasons}.

In section \ref{sect:highdensity}, the thermal Jeans mass, $M_J$,
was shown to have a dependence on $\left(\langle a^2D\rangle
/Z^\alpha\right)^{3/4}n^{1/4}$ for grain radius $a$, dust-to-gas
particle number ratio $D$, density $n$, and metallicity dependence
$Z^\alpha$ that appears in the gas cooling expression. We noted how
this would not depend much on metallicity as collisional processes
bias the average value $\langle a^2D\rangle$ toward small grains
while metallicity, containing the average of $a^3D$, has a greater
weight for large grains.  However, when the relative dust mass gets
low, $\langle a^2D\rangle$ starts to drop. If we consider a Mathis
et al. (1977) grain size distribution $dn_g/da=n_a a^{-3.5}$, and a
relative dust mass abundance proportional to the metallicity, $Z$,
then $Z/Z_\odot=\left(a_{max}/a_{max,\odot}\right)^{1/2}\xi
A/A_\odot$ for $\xi=n_a/n_{a\odot}$ and
$A=1-\left(a_{min}/a_{max}\right)^{1/2}$. The minimum and maximum
grain sizes are denoted by subscripts. Solving for $A$ and
integrating over the grain size distribution again, we get
\begin{equation}
{{\langle a^2D\rangle}\over{\langle a^2D\rangle_\odot}}=
{{Z/Z_\odot}\over{\left(Z/Z_\odot\right)\left(A_\odot/\xi\right)+
(a_{min,\odot}/a_{max,\odot}})^{1/2}}\label{eq:13}\end{equation} for
constant $a_{min}$. For normal metallicities, when
$Z/Z_\odot>\left(\xi/A_\odot\right)
\left(a_{min,\odot}/a_{max,\odot}\right)^{1/2}$, $\langle
a^2D\rangle\propto n_a$ and is nearly independent of $Z/Z_\odot$, as
mentioned above. For very low metallicities,
$Z/Z_\odot<\left(\xi/A_\odot\right)\left(a_{min,\odot}/a_{max,\odot}\right)^{1/2}$,
$\langle a^2D\rangle$ decreases with $Z/Z_\odot$. Unless $\alpha$ in
the cooling rate is larger than 1, i.e., the molecular gas cooling
rate depends sensitively on metallicity, $\langle a^2D
\rangle/Z^\alpha$ decreases at very low metallicity. This lowers
$M_J$, which is opposite to the effect observed in the early
universe. Thus, lower gas cooling at small $Z$ does not necessarily
produce larger $M_J$ at the grain-gas coupling density.

In section \ref{sect:ambient}, ambient ISM conditions were considered
rather than dense cores. $M_J$ was shown to be insensitive to density
and radiation field as long as the Kennicutt star formation law is
satisfied. If the star formation rate is much higher for a given
density than it is locally, then the dust temperature can be higher for
a given density, and $M_J$ higher. This requires a higher efficiency of
star formation. Such deviations from the Kennicutt relation have been
suggested for ultraluminous infrared galaxies (Graci\'a-Carpio et al.
2008). Thus, higher efficiencies could have caused the observed $M_c$
increase. The results in section \ref{sect:ambient} also depend on
equation 3, which is the local relation between grain temperature and
radiation field. If the metallicity is much lower and the grains are
systematically smaller and hotter, $M_J$ will be higher for a given
relation between radiation field and density. Observations of 66
starburst galaxies by Engelbracht et al. (2008) showed $T_d\propto
Z^{-0.2}$ down to $Z/Z_\odot \sim 0.1$ (their Fig. 5), although $T_d$
appears to drop with decreasing $Z/Z_\odot$ below that. Setting
M$_J\propto T_d^{3/2}$ for this Engelbracht et al. (2008) relation, we
get $M_J \propto Z^{-0.3}$ if the Kennicutt relation still holds. Thus
decreases in metallicity could correspond to increases in $M_J$ because
of increased ambient grain temperatures at the same density and
radiation field. It is not clear if this trend should continue for
$Z/Z_\odot < 0.1$, where the Engelbracht et al. (2008) power-law
relation stops. Equations \ref{eq:dust} and \ref{eq:13} suggest that it
should. In that case, the extremely metal-poor Milky Way halo stars
investigated by Komiya et al.\ (2007), which have [Fe/H]$<-2.5$, would
have higher $M_J$ by more than a factor of $10^{2.5*0.3}=5.6$, which is
about the $M_c$ shift that they observe.

Section \ref{sect:compressed} considers $M_J$ in highly ionized
regions. The same dependence of $M_J$ on $Z$ should result if the
grains become systematically hotter for lower metallicity with a given
radiation field. A more prominent effect in strong radiation fields may
be the complete ionization of pre-stellar globules, even if M$_J$
itself does not change much. Equation \ref{eq:ionize} indicates that
sufficiently small globules are destroyed by ionization before they
collapse to a star. We were inconclusive about the importance of this
effect because it depends sensitively on the globule density and on the
column density of the surrounding cloud. If the cosmological increase
in $M_J$ depends more on star formation rate and star formation density
than it does on metallicity, then the ionization of small prestellar
globules could be the primary explanation for the lack of low mass
stars. For example, if $M_c$ is high in starbursting mergers during the
formation of an elliptical galaxy, and the metallicity is somewhat
solar, then the ionization of pre-stellar globules could be an
important factor.  Metallicity could still play a role because dust
affects the scattering of ionizing photons.

For all of these scenarios, it is difficult to understand why Milky Way
globular clusters, with metallicities [Fe/H]$\sim-1.5$ (Binney \&
Tremaine 1987), have characteristic masses that appear similar to those
of modern metal-rich super star clusters. According to the $M_J \propto
Z^{-0.3}$ relation above, $M_J$ should be larger by a factor of $\sim3$
in globular clusters if evaporation of pre-stellar cores is not
effective, and larger still if it is affected by pre-stellar
evaporation. A factor of three may be too small to notice after a
Hubble time of mass segregation and evaporation in these clusters.
However, such a shift is consistent with that suggested by Chabrier
(2003) for the Milky Way bulge and globular clusters, as noted in
section 1. More observations of the characteristic stellar mass as a
function of metallicity or average grain size are needed.

\section{Conclusions}
\label{sect:conclusions}

The characteristic mass for star formation, which appears as a plateau
in the IMF, is nearly constant over a wide range of conditions, ranging
from the outer part of the Milky Way, to dense molecular cores, to
super star clusters. This constancy is explained as the result of very
weak dependencies for the thermal Jean mass $M_J$ on density,
metallicity, and radiation field in three fundamental environments:
dense cloud cores, the ambient ISM, and the vicinity of highly ionizing
super star clusters. Dense cores give a constant $M_J$ because of the
way the molecular cooling rate scales with density and temperature. The
ambient ISM has a constant $M_J$ because of star formation feedback
with the Kennicutt-Schmidt law. Dense ionizing clusters have constant
$M_J$ because of interdependencies between the pressure and radiation
field, which is coupled to the dust temperature. In all cases, $M_J$
varies by only a factor of $\sim4$ or less in the most extreme
situations.

These three cases are representative of a wide range of conditions for
star formation. The values of $M_J$ are about the same for each, and in
all cases, this value may be identified with the characteristic mass of
star formation, $M_c$.   The fact that $M_J$ is so invariant with
environment, even for different ways of looking at environment, seems
to explain the constancy of the characteristic mass in the IMF.

Observations of real increases in $M_J$ under cosmological
conditions were summarized. The environment for star formation is
not well understood in this case, but there are some fundamental
ways in which the physical parameters are expected to change in
extreme conditions. One possibility is that the efficiency of star
formation increases, so the radiation field is stronger for a given
density and pressure. Another possibility is that the minimum grain
temperature increases for a given radiation field as a result of a
decreasing maximum grain size at low metallicities. A third
possibility is that high radiation densities in super starburst
conditions evaporate low-mass prestellar cores. Observations of any
correlations between the lower part of the IMF and environmental
conditions at high redshift would be useful in understanding the
characteristic mass.

\acknowledgments

We thank Frank Israel, Susanne H{\"u}ttemeister,  Marco Spaans and an
anonymous referee for valuable comments. This study was triggered by
discussions at the Kavli Institute for Theoretical Physics in Santa
Barbara and thus supported in part by the National Science Foundation
under Grant No. PHY05-51164. Partial support is also provided by the
German Science Foundation (DFG) via the priority program SFB 439
"Galaxies in the Early Universe" and by a grant to C.D.W. from the
Natural Science and Engineering Research Council of Canada (NSERC).

\begin{deluxetable}{lccl}
\tabletypesize{\scriptsize} \tablewidth{0pt} \tablecaption{Mass
Range for the IMF Plateau\label{table1}} \tablehead{\colhead{Region}
&\colhead{Lower Mass ($M_\odot$)} &\colhead{Upper Mass ($M_\odot$)}
&\colhead{Reference}} \startdata
Taurus       & 0.1  &0.8   &Luhman 2004\\
$\rho$ Oph   & $\le$ 0.1  &0.5   &Luhman \& Rieke 1999\\
IC 348       &0.1  &1     &Luhman et al.\ 2003\\
Cam I        &0.1  &1     &Luhman 2007\\
Trapezium    &0.1  &0.6   &Muench et al.\ 2002\\
Pleiades     &0.2 & 0.7     &Bouvier et al.\ 1998\\
Upper Sco OB & $\le$0.1    &0.6   &Preibisch et al.\ 2002\\
M35          &0.2  &0.8   &Barrado y Navascu\'es et al.\ 2001\\
Local Field  &0.2  &0.5   &Scalo 1986\\
"            &0.1  &1     &Rana 1987\\
"            &0.1  & 0.5     &Kroupa, Tout, \& Gilmore 1993\\
12 GCs\tablenotemark{a}  &0.14  &0.69  &Parsece \& de Marchi 2000\\
M4 (GC)\tablenotemark{b} &0.14  &0.69  &de Marchi et al.\ 2004\\
M15 (GC)\tablenotemark{c}&0.16 &0.57  &Pasquali et al.\ 2004\\
MW bulge     & ... & 0.5-0.7  &Holtzman et al.\ 1998\\
"            & $\le$0.15    &  0.5    &Zoccali et al.\ 2000\\
\enddata
\tablenotetext{a}{For these Milky Way globular clusters, the initial
mass function for $M<0.75\;M_\odot$ was found to be consistent with a
log normal having a peak mass of $M_p=0.33\pm0.03$ ($\log M_c=-0.5$)
and a dispersion in the log-mass of $\sigma=0.34\pm0.04$. The lower and
upper limits of the plateau were taken to be the one-$\sigma$ points on
either side of the peak.} \tablenotetext{b}{De Marchi et al.\ (2004)
fit the mass function for M4 to a tapered power law,
$M^{-\alpha}\left[1-e^{\left(-M/M_p\right)^{-\beta}}\right]$ and find
values of $\alpha$, $M_p$, and $\beta$ to be essentially the same as
for the 12 globular clusters in Paresce \& DeMarchi (2000). Thus we
take the same range for the plateau.} \tablenotetext{c}{Pasquali et
al.\ (2004) fit a log-normal mass function to this globular cluster
with peak mass $M_p=0.3$ and dispersion $\sigma=0.28$. The mass limits
in the table are one-$\sigma$ points around the peak.}
\end{deluxetable}

{}

\newpage

\begin{thebibliography}{}

\bibitem[]{917} Abel, T., Bryan, G.~L., \& Norman, M.~L. 2002, Science, 295, 93

\bibitem[]{919} Adams, F. C., \& Fatuzzo, M. 1996, ApJ, 464, 256

\bibitem[]{921} Alves, J., Lombardi, M., \& Lada, C.J. 2007, A\&A,
462, L17

\bibitem[]{924} Andersen, M., \& Zinnecker, H. 2003, in Star Formation Through Time, ASP, 297,
eds. E. Perez, R.M. Gonzalez Delgado, \&  G. Tenorio-Tagle, p. 115

\bibitem[Andersen et al.(2005)]{2005IAUS..227..285A} Andersen, M., Meyer,
M.~R., Greissl, J., Oppenheimer, B.~D., Kenworthy, M.~A., McCarthy, D.~W.,
\& Zinnecker, H.\ 2005, in Massive Star Birth: A Crossroads of Astrophysics, IAU Symposium 227, eds.\ R.\ Cesaroni, M.\ Felli, E.\ Churchwell, M.\  Walmsley (Cambridge University Press), p.\ 285

\bibitem[]{931} Ballero, S. K., Kroupa, P., \& Matteucci, F. 2007, A\&A,
467, 117

\bibitem[]{934} Barrado y Navascu\'es, D., Stauffer, J.R., Bouvier,
J., \& Mart\'in, E.L. 2001, ApJ, 546, 1006

\bibitem[]{937} Bastian, N., \& Goodwin, S. P. 2006, MNRAS, 369, L9

\bibitem[]{939} Bate, M.~R., Bonnell, I.~A., \& Bromm, V.\ 2003, \mnras, 339, 577

\bibitem[]{941} Bate, M.R., \& Bonnell, I.A. 2005, MNRAS, 356, 1201

\bibitem[]{943} Beers, T.~C., \& Christlieb, N. 2005, ARA\&A, 43, 531

\bibitem[]{945} Bertoldi, F., \& McKee, C.F. 1990, ApJ, 354, 529

\bibitem[]{947} Beuther, H., \& Schilke, P. 2004, Science, 303, 1167

\bibitem[]{949} Binney, J., \& Tremaine, S. 1987, Galactic Dynamics,
Princeton University Press.

\bibitem[]{954} Boissier, S., Prantzos, N., Boselli, A., \& Gavazzi,
G. 2003, MNRAS, 346, 1215

\bibitem[]{957} Bonnell, I., Clarke, C.J., \& Bate, M.R. 2006, MNRAS,
368, 1296

\bibitem[]{960} Bonnell, I.A., Larson, R.B., \& Zinnecker, H. 2007,
Protostars and Planets, ed. B. Reipurth, et al.\ (Tucson: Univ. of
Arizona), p. 149

\bibitem[]{964} Bouvier, J., Stauffer, J. R., Martin, E. L., Barrado y
Navascu\'es, D., Wallace, B., \& Bejar, V. J. S. 1998, A\&A, 336,
490

\bibitem[]{968} Bromm, V., Ferrara, A., Coppi, P.~S., \& Larson, R.~B. 2001, MNRAS, 328, 969

\bibitem[]{970} Bromm, V., Coppi, P.S., \& Larson, R.B. 2002, ApJ,
564, 23

\bibitem[]{973} Chabrier, G. 2003, PASP, 115, 763

\bibitem[]{975} Charlot, S., Ferrari, F., Matthews, G. J., \& Silk, J.
1993, ApJ, 419, L57

\bibitem[]{978} Chiosi, C. 2000, A\&A, 364, 423

\bibitem[]{980} Christlieb, N., Bessell, M. S., Beers, T. C., Gustafsson, B., Korn, A., Barklem, P. S., Karlsson, T., Mizuno-Wiedner, M., \& Rossi, S.  2002, Nature, 419, 904

\bibitem[]{982} Clark, P.C. \& Bonnell, I.A. 2005, MNRAS, 361, 2

\bibitem[]{984} Clark, P.~C., Glover, S.~C.~O., \& Klessen, R.~S. 2008, ApJ, 672, 757

\bibitem[]{986} Clark, P.\ C., Klessen, R.\ S., \& Bonnell, I.\ A. 2007, MNRAS, 379, 57

\bibitem[]{988} Crapsi, A., Caselli, P., Walmsley, C.M., \& Tafalla,
M. 2007, A\&A, 470, 221

\bibitem[]{991} Dav\'e, R. 2008, MNRAS, in press

\bibitem[]{993} de Grijs, R., \& Parmentier, G. 2007, ChJAA, 7, 155

\bibitem[]{995} de Marchi, G., Leibundgut, B., Paresce, F., \& Pulone,
L. 1999, A\&A, 343, 9

\bibitem[]{998} de Marchi, G., Paresce, F., \& Pulone, L. 2000, ApJ,
530, 342

\bibitem[]{1001} de Marchi, G., Paresce, F., Straniero, O., \& Moroni,
P. 2004, A\&A, 415, 971

\bibitem[]{1004} de Marchi, G., Paresce, F., \& Portegies Zwart, S.
2004, astroph/0409601

\bibitem[]{1007} de Marchi, G., Pulone, L. \& Paresce, F. 2006, A\&A,
449, 161

\bibitem[]{1010} de Marchi, G., \& Pulone, L. 2007, A\&A, 467, 107

\bibitem[]{1012} Elmegreen, B.\ G., 1993, ApJ, 419, L29

\bibitem[]{1014} Elmegreen, B.G. 2002, ApJ, 577, 206

\bibitem[]{1016} Elmegreen, B.G. 2005,  in Starbursts: from 30 Doradus
to Lyman Break Galaxies,  Dordrecht: Springer, ed. R. de Grijs \&
R.M. Gonzalez Delgado, ApSpSci Library, Vol. 329, p.57

\bibitem[]{1020} Elmegreen, B.G. 2006, ApJ, 648, 572

\bibitem[]{1022} Elmegreen, B.\ G. 2007, ApJ, 668, 1064

\bibitem[]{1024} Elmegreen, B.G., \& Efremov, Y.N. 1996, ApJ, 466, 802

\bibitem[]{1026} Engelbracht, C.W., Rieke, G.H., Gordon, K.D., Smith,
J.-D., T., Werner, M.W., Moustakas, J., Willmer, C.N.A., \& Vanzi,
L. 2008, astroph/0801.1700

\bibitem[]{1030} Fardal, M.A., Katz, N., Weinberg, D.H., \& Dav\'e,
R. 2007, MNRAS, 379, 985

\bibitem[]{1033} Feltzing, S., Gilmore, G., \& Wyse, R.F.G. 1999,
ApJL, 516, 17

\bibitem[]{1036} Field, G.\ B., Saslaw, W.\ C. 1965, ApJ, 142, 568

\bibitem[]{1038} Fleck, R.\ C., 1982, MNRAS,  201, 551

\bibitem[]{1040} Frebel, A., Johnson, J.~L., \& Bromm, V. 2007, MNRAS, 380, L40

\bibitem[]{1042} Graci\'a-Carpio, J., Garc\'ia-Burillo, S., Planesas, P.,
Fuente, A., \& Usero, A. 2008, A\&A, 479, 703

\bibitem[]{1045} Gould, A., Bahcall,
J.~N., \& Flynn, C.\ 1997, ApJ, 482, 913

\bibitem[]{1048} Heger, A., \& Woosley, S.~E. 2002, ApJ, 567, 532

\bibitem[]{1050} Hollenbach, D.J. 1989, in Interstellar Dust, eds.
L.J. Allamandola \& A. G. G. M. Tielens. IAUSymp. 135, Dordrecht:
Kluwer, p.227

\bibitem[]{1054} Holtzman, J. A., Watson, A. M., Baum, W.A.,
Grillmair, C.J., Groth, E.J., Light, R.M., Lynds, R., O'Neil, E.J.,
Jr. 1998, AJ, 115, 1946

\bibitem[]{1058} Ikeda, N., Sunada, K., \& Kitamura, Y. 2007,
ApJ, 665, 1194

\bibitem[]{1061} Jappsen, A.-K., Klessen, R.S., Larson, R.B., Li, Y.,
\& MacLow M.-M. 2005, A\&A, 435, 611

\bibitem[]{1064} Jappsen, A.- K., Klessen, R.\ S., Glover, S.\ C.\ O., \& Mac~Low, M.-M. 2007, ApJ, submitted (arXiv:0709.3530)

\bibitem[]{1066} Johnstone, D., Wilson, C. D., \& Moriarty-Schieven,
G., et al. 2000, ApJ, 545, 327

\bibitem[]{1069} Johnstone, D., Fich, M., Mitchell, G. F., \&
Moriarty-Schieven, G. 2001, ApJ, 559, 307

\bibitem[]{1072} Kennicutt, R.C., Jr. 1998, ApJ, 498, 541

\bibitem[]{1074} Kennicutt, R.C., Jr., Calzetti, D., Walter,
F., Helou, G., Hollenbach, D.J., Armus, L., Bendo, G., Dale, D.A.,
Draine, B.T., Engelbracht, C.W., Gordon, K.D., Prescott, M.K.M.,
Regan, M.W., Thornley, M.D., Bot, C., Brinks, E., de Blok, E., de
Mello, D., Meyer, M., Moustakas, J., Murphy, E.J., Sheth, K., \&
Smith, J. D. T. 2007, astro-ph/0708.0922

\bibitem[]{1081} Koch, A., Grebel, E.K., Odenkirchen, M.,
Mart\'inez-Delgado, D. \& Caldwell, J. A. R. 2004, AJ, 128, 2274

\bibitem[]{1084} Koen, C. 2006, MNRAS, 365, 590

\bibitem[]{1086} Komiya, Y., Suda, T.,
Minaguchi, H., Shigeyama, T., Aoki, W., \& Fujimoto, M.~Y.\ 2007,
ApJ, 658, 367

\bibitem[]{1090} Klessen, R.\ S. 2001, ApJ, 556, 837

\bibitem[]{1092} Klessen, R.\ S., Burkert, A., \& Bate, M.\ R. 1998, \apj, {\bf 501}, L205

\bibitem[]{1094} Klessen, R.\ S., \& Burkert, A.  2000,  ApJS,  128, 287

\bibitem[]{1096} Klessen, R.\ S., \& Burkert, A.  2001,  ApJ,  549, 386

\bibitem[]{1098} Klessen, R.\ S., Spaans, M., Jappsen, A.-K. 2007, MNRAS, 374, L29

\bibitem[]{1100} Kroupa, P. 2002, Science, 295, 82

\bibitem[]{1102} Krumholz, M.R., \& McKee, C.F. 2005, ApJ, 630, 250

\bibitem[]{1104}Komiya, Y., Suda, T., Minaguchi, H., Shigeyama, T.,
Aoki, W., \& Fujimoto, M.~Y. 2007, ApJ, 658, 367

\bibitem[]{1107} Lada, C.\ J., Muench, A.\ A., Rathborne, J., Alves, J.\ F., \& Lombardi, M. 2008, ApJ, 672, 410

\bibitem[]{1109} Larson, R.B. 1978, MNRAS, 184, 69

\bibitem[]{1111} Larson, R.\ B., 1981, MNRAS, 194, 809

\bibitem[]{1113} Larson, R. B. 1982, MNRAS, 200, 159

\bibitem[]{1115} Larson, R.\ B. 1985, MNRAS, 214, 379

\bibitem[]{1117} Larson, R.B. 2005, MNRAS, 359, 211

\bibitem[]{1119} Larson, R.B. 2007, astroph/0701733

\bibitem[]{1121} Li, Y., Klessen, R.\ S., \& Mac Low, M.-M. 2003, ApJ, 592, 975

\bibitem[]{1123} Li, D., Velusamy, T., Goldsmith, P. F., \& Langer,
W.D. 2007, ApJ, 655, 351

\bibitem[]{1126} Loewenstein, M., \& Mushotsky, R.F. 1996, ApJ, 466,
695

\bibitem[]{1129} Lucatello, S., Tsangarides, S., Beers, T. C., Carretta, E., Gratton, R. G., Ryan, S. G. 2005, ApJ, 625, 825

\bibitem[]{1131} Luhman, K.L. 2004, ApJ, 617, 1216

\bibitem[]{1135} Luhman, K.L. 2007, ApJS, 173, 104

\bibitem[]{1137} Luhman, K.L. \& Rieke, G.H., 1999, ApJ, 525, 440

\bibitem[]{1141} Luhman, K.L., Stauffer, J.R., Muench, A.A., Rieke,
G.H., Lada, E.A., Bouvier, J., \& Lada, C.J. 2003, ApJ, 593, 1093

\bibitem[]{1144} Luna, A., Bronfman, L., Carrasco, L., \& May, J. 2006,
ApJ, 641, 938

\bibitem[]{1147} Mac Low, M.-M. \& Klessen, R.~S. 2004, Rev.~Mod.~Phys., 76, 125

\bibitem[]{1149} Maness, H., Martins, F., Trippe, S., Genzel, R.,
Graham, J. R., Sheehy, C., Salaris, M., Gillessen, S., Alexander,
T., Paumard, T., Ott, T., Abuter, R., \& Eisenhauer, F. 2007, ApJ,
669, 1024

\bibitem[]{1154} Martel, H., Evans,
N.~J., II, \& Shapiro, P.~R.\ 2006, \apjs, 163, 122

\bibitem[]{1157} Martin, C.L., \& Kennicutt, R.C. 2001, ApJ, 555, 301

\bibitem[]{1159} Massi, F., de Luca, M., Elia, D., Giannini, T., Lorenzetti, D., \& Nisini,
B. 2007, A\&A, 466, 1013

\bibitem[]{1162} Masunaga, H. \& Inutsuka, S.-I. 1999, ApJ, 510, 822

\bibitem[]{1164} Masunaga, H. \& Inutsuka, S.-I. 2000, ApJ, 531, 350

\bibitem[]{1166} Mathis, J. S., Rumpl, W., \& Nordsieck, K. H. 1977,
ApJ, 217, 425

\bibitem[]{1169} Miller G. E. \& Scalo J. M., 1979, ApJS, 41, 513

\bibitem[]{1171} Moraux, E., Bouvier, J., Stauffer, J. R., Barrado y Navascués,
D., \&Cuillandre, J.-C. 2007, A\&A, 471, 499

\bibitem[]{1174} Moretti, A., Portinari, L., \& Chiosi, C. 2003,
A\&A, 408, 431

\bibitem[]{1177} Motte, F., Andr\'e, P., \& Neri, R. 1998, A\&A, 336,
150

\bibitem[]{1180} Motte, F., Andr\'e, P., Ward-Thompson, D., \&
Bontemps, S. 2001, A\&A, 372, L41

\bibitem[]{1183} Muench, A.A., Lada, E.A., Lada, C.J., \& Alves, J.
2002, ApJ, 573, 366

\bibitem[]{1186} Murray, S.\ D., and D.\ N.\ C.\ Lin,  1996,  ApJ,  467,  728

\bibitem[]{1188} Nagashima, M., Lacey, C. G., Baugh, C.M., Frenk,
C.S., \& Cole, S. 2005a, MNRAS, 363, 1247

\bibitem[]{1191} Nagashima, M., Lacey, C. G., Okamoto, T., Baugh,
C.M., Frenk, C.S., \& Cole, S. 2005b, MNRAS, 363, L31

\bibitem[]{1194} Nakano, T., Hasegawa, T., \& Norman, C. 1995, ApJ,
450, 183

\bibitem[]{1197} Nayakshin, S., \& Sunyaev, R. 2005, MNRAS, 364, L23.


\bibitem[]{1200} Neufeld, D. A., Lepp, S., \& Melnick, G. J. 1995, ApJS, 100, 132

\bibitem[]{1202} Nutter, D., \& Ward-Thompson, D. 2007, MNRAS, 374, 1413

\bibitem[]{1204} Oey, M. S., \& Clarke, C. J. 2005, ApJ, 620, L43

\bibitem[]{1206} Omukai, K., Tsuribe, T., Schneider, R., \& Ferrara, A. 2005, ApJ, 626, 627

\bibitem[]{1208} O'Shea, B.~W., \& Norman, M.~L. 2007, ApJ, 654, 66

\bibitem[]{1210} Padoan, P., 1995, MNRAS, 277, 377

\bibitem[]{1212} Padoan, P., \& Nordlund, \AA. 2002, ApJ, 576, 870

\bibitem[]{1214} Padoan, P., Nordlund, \AA., \& Jones, B.\ J.\ T. 1997, MNRAS, 288, 145

\bibitem[]{1216} Paresce, F. \& de Marchi, G. 2000, ApJ, 534, 870

\bibitem[]{1218} Pasquali, A., de Marchi, G., Pulone, L., \& Brigas,
M.S. 2004, A\&A, 428, 469

\bibitem[]{1221} Paumard, T., Genzel, R., Martins, F., Nayakshin, S.,
Beloborodov, A. M., Levin, Y., Trippe, S., Eisenhauer, F., Ott, T.,
Gillessen, S., Abuter, R., Cuadra, J., Alexander, T., \& Sternberg,
A. 2006, ApJ, 643, 1011

\bibitem[]{1226} Pipino, A., \& Matteucci, F. 2004, MNRAS, 347, 968

\bibitem[]{1228} Portinari, L., Moretti, A., Chiosi, C., \&
Sommer-Larsen, J. 2004, ApJ, 604, 579

\bibitem[]{1231} Preibisch, T., Brown, A.G.A., Bridges, T., Guenther,
E. \& Zinnecker, H. 2002, AJ, 124, 404

\bibitem[]{1234} Rana, N.C. 1987, A\&A, 184, 104

\bibitem[]{1236} Rees, M.J. 1976, MNRAS, 176, 483

\bibitem[]{1238} Reid, N. 1987, MNRAS, 225, 873

\bibitem[]{1240} Renzini, A., Ciotti, L., D'Ercole, A., \&
Pellegrini, S. 1993, ApJ, 419, 52

\bibitem[]{1243} Robertson, B.E., \& Kravtsov, A.V. 2007,
astroph/0710.2102

\bibitem[]{1246} Romeo, A.D., Sommer-Larsen, J., Portinari, L., \&
Antonuccio-Delogu, V. 2006, MNRAS, 371, 548

\bibitem[]{1249} Ryan, S.~G., Aoki, W., Norris, J.~E., \& Beers, T.~C. 2005, ApJ, 635, 349

\bibitem[]{1251} Salpeter, E. 1955, ApJ, 121, 161

\bibitem[]{1253} Sanders, D.B., Solomon, P.M., \& Scoville, N.Z. 1984,
ApJ, 276, 182

\bibitem[]{1256} Scalo, J.M. 1986, Fund.Cos.Phys, 11, 1

\bibitem[]{1258} Scalo, J., 1990, in Astrophys.\ \& Space Lib., 160, 125

\bibitem[]{1260} Scalo, J.M. 1998, in The Stellar Initial Mass
Function, ed. G. Gilmore, I. Parry, \& S. Ryan, Cambridge: Cambridge
University Press, p. 201

\bibitem[]{1264} Schaye, J. 2004, ApJ, 609, 667

\bibitem[]{1266} Schneider, R., Ferrara, A., Natarajan, P. \& Omukai, K. 2002, ApJ, 571, 30

\bibitem[]{1268} Schneider, R., Omukai, K., Inoue, A.~K., \& Ferrara, A. 2006, MNRAS, 369, 1437

\bibitem[]{1270} Shu, F.H., Adams, F.C., \& Lizano, S. 1987, ARAA, 25,
23

\bibitem[]{1273} Silk, J. 1977, ApJ, 214, 718

\bibitem[]{1275} Sirianni, M., Nota,
A., Leitherer, C., De Marchi, G., \& Clampin, M.\ 2000, ApJ, 533,
203

\bibitem[]{1279} Stanke, T., Smith, M.D., Gredel, R., \& Khanzadyan, T.
2006, A\&A, 447, 609

\bibitem[]{1282} Stolte, A., Brandner, W., Grebel, E.K., Lenzen, R. \&
Lagrange, A.-M. 2005, ApJ, 628, L113

\bibitem[]{1285} Testi, L., \& Sargent, A. I. 1998, ApJ, 508, L91

\bibitem[]{1287} Tielens, A.G.G.M. 2005, The Physics and Chemistry of
the Interstellar Medium, Cambridge University Press.

\bibitem[]{1290} Tohline, J. E. 1980, ApJ, 239, 417

\bibitem[]{1292} Tornatore, L., Borgani, S., Matteucci, F., Recchi,
S., \& Tozzi, P. 2004, MNRAS, 349, L19

\bibitem[]{1295} Tumlinson, J. 2006, ApJ, 641, 1

\bibitem[]{1297} Vacca, W. D., Garmany, C. D., \& Shull, J. M. 1996, ApJ, 460, 914

\bibitem[]{1299} van Dokkum, P. 2008, ApJ, 674, 29

\bibitem[]{1301} V{\'a}zquez-Semadeni, E., Ballesteros-Paredes, J., \& Klessen, R.\ S. 2003, ApJ, 585, L131

\bibitem[]{1303} Vorobyov, E.I. 2003, A\&A, 407, 913

\bibitem[]{1305} Walsh, A.J., Myers, P.C., Di Francesco,
J., Mohanty, S., Bourke, T.L., Gutermuth, R., \& Wilner, D. 2007,
ApJ, 655, 958

\bibitem[]{1309} Wang, B., \& Silk, J. 1993, ApJ, 406, 580

\bibitem[]{1311} Weidner, C., \& Kroupa, P. 2004, MNRAS, 348, 187

\bibitem[]{1313} Whitworth, A.P., Boffin, H.M.J., \& Francis, N. 1998,
MNRAS, 299, 554

\bibitem[]{1316} Yang, Y., Park, H.S., Lee, M.G. \& Lee, S.G. 2002,
JKAS, 35, 131

\bibitem[]{1319} Yasui, C., Kobayashi, N., Tokunaga, A.T., Terada, H., \& Saito,
M. 2008, ApJ, 675, 443

\bibitem[]{1322} Yoshida, N., Omukai, K., Hernquist, L., \& Abel, T. 2006, ApJ, 652, 6

\bibitem[]{1324} Yoshii, Y., \& Saio, H 1985, ApJ, 295, 521

\bibitem[]{1326} Zoccali, M., Cassisi, S., Frogel, J.A., Gould, A.,
Ortolani, S., Renzini, A., Rich, R. M., \& Stephens, A.W. 2000, ApJ,
530, 418

\end{thebibliography}
\end{document}